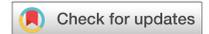

OPEN

# Artificial magnetic conductor backed dual-mode sectoral cylindrical DRA for off-body biomedical telemetry

Nayab Gogosh[1,2], Sohail Khalid[1], Bilal Tariq Malik[3] & Slawomir Koziel[3,4]✉

This research investigates the potential of a sectoral Cylindrical Dielectric Resonator Antenna (CDRA) for biomedical telemetry. CDRAs are known for their low-loss, ruggedness, and stability, but their limited bandwidth and size make them unsuitable for wearable devices. The research addresses these limitations by proposing a dual-mode antenna that operates in $EH_{110}$ and $TE_{210}$ modes. The sectoral CDRA is a quarter segment with Perfect Electric Conductor boundaries, reducing its size by a factor of four. Mathematical derivations of the field components for both modes are derived to support the design. To minimize specific absorption rate (SAR), an Artificial Magnetic Conductor (AMC) surface is applied to the antenna's backside, enhancing compatibility with the transverse electric modes. The antenna achieves a bandwidth of 0.7 GHz (5.2–5.9 GHz), suitable for biomedical applications, with a measured peak gain of 7.9 dBi and a SAR of 1.24 W/kg when applied to a human arm.

Biomedical telemetry is an emerging interdisciplinary domain where body vitals are transmitted wirelessly for detailed observation and analysis[1]. Significant advancements in semiconductor technology have led to numerous breakthroughs, providing highly compact and energy-efficient electronic devices for processing and communication purposes. Nevertheless, the large size of the antenna continues to present challenges in achieving compactness for sensing devices, as the laws of physics dictate the limitations in miniaturizing microwave passive components. Furthermore, the restrictions imposed by antenna losses and specific absorption rates of these radiating devices hinder the development of compact and high-performance devices[1,2].

The use of biomedical devices on the body is widely accepted due to their noninvasive nature and the ease of their removal and replacement[1,3]. In recent years, several antennas have been developed for telemetric applications on the human body[4–6]. These antennas have mostly developed from printed antenna technology with ohmic losses and gain limitations. Dielectric Resonator Antennas (DRAs) are a viable substitute for printed antennas, as they provide excellent performance while eliminating conductive losses[7–10]. DRA can be realized in different geometries such as square, hemispherical, conical, spherical, and cylindrical. However, cylindrical DRAs (CDRAs) offer specific benefits such as mode symmetry, transverse electric or magnetic modes, and the ability to achieve hybrid modes with construction and placement convenience[11–15]. Limited research has been reported in the literature on the use of DRAs for wearable applications[16–23]. A button-shaped CDRA has been presented in[16]. The antenna is worn on the shoulder like a button and operates at a frequency of 5.8 GHz. It has a gain of 5.4 dBi and a narrow bandwidth of 7%. Another antenna design has been introduced in[17], which combines a circularly polarized dielectric resonator antenna with a loop antenna in its center. The antenna operates at a frequency of 2.4 GHz and has a limited bandwidth of 4.95%. The antenna exhibits a low gain of − 0.1 dBi. A recent study has introduced a compact rectangular DRA constructed using Alumina, which produces $TE^{x}_{\delta\,11}$ and $TE^{y}_{1\delta\,1}$ modes[18]. The antenna operates in the frequency range of 4.72 to 9.39 GHz, providing 66% bandwidth. It achieves a maximum gain of 7.5 dBi. The irregularity of the DRA form factor complicates its practical realization. The authors in[19] have introduced a concept for annular conical DRA tailored for body area networks. The antenna design provides a bandwidth of 38% within the frequency range of 3.4 GHz to 5 GHz.

[1]Electrical and Computer Engineering Department, Riphah International University, Islamabad, Pakistan. [2]Electrical Engineering Department, COMSATS University Islamabad, Park Road, Islamabad, Pakistan. [3]Faculty of Electronics, Telecommunications, and Informatics, Gdansk University of Technology, Gdansk 80-233, Poland. [4]Engineering Optimization and Modeling Center, Reykjavik University, 101 Reykjavik, Iceland. ✉email: koziel@ru.is

The DRA is fed through a capacitive-loaded monopole antenna, which is probed at its center to generate the $TM_{01\delta}$ resonant mode. The conical DRA has a height of 12.5 mm and the radius of the antenna, including the substrate, is 30 mm. A square DRA has been reported for body area network applications in[20]. The antenna is fed through an H-shaped metal strip connected to one of its sides. The antenna has an operating bandwidth of 9.6% from 7.47 to 8.25 GHz. The antenna provides a gain of 5 dBi with a large thickness of 15 mm. A recent study has introduced a wristwatch-like DRA that has been applied to forearms and ankles in the 2.5 GHz frequency band[21]. The antenna consists of a slightly curved rectangular DR attached to a conformal substrate. The antenna feed is implemented as a conformal strip that excites the DR from its side. The antenna produces a $TM_{11\delta}$ mode with a maximum measured gain of 4.6 dBi and a thickness of 9.44 mm. Another study has introduced a wearable textile DRA[22]. The antenna is transformed into a bowtie shape by selectively removing parts of the CDRA. The antenna has a height of 12.7 mm and a radius of 10 mm. It is connected to a microstrip transmission line for feeding. The antenna provides a wide frequency range of 4.3 GHz to 6.4 GHz with a bandwidth of 40%. Several wearable button-shape antennas have also been reported for wearable applications[23–27]. A circularly polarized FR4 button antenna, fed by a probe, has been proposed for the ISM and U-NII frequency bands[23]. The antenna provides a gain of 3.9 dBi and has an axial ratio bandwidth of 7%. The antenna is fixed 12 mm above the conductive textile substrate. In[24], a dual-band antenna has been reported that operates in both the 2.4 and 5.8 GHz frequency bands with the respective gains of -0.6 dBi and 4.3 dBi.

It can be noted that the DRAs designed for on-body applications are either bulky, particularly in terms of thickness, or exhibit low gain. Since compactness is an important aspect of biomedical antennas and to achieve miniaturization in CDRA, sectoral antennas have been explored in the literature[28,29]. The investigation of the sector CDRA concept has predominantly focused on TM modes due to their straightforward excitation process. This study is the first to investigate the TE and hybrid modes and propose a direct feeding mechanism to achieve a dual-mode response in sectoral CDRAs. In addition, the mathematical derivation of the field components for these modes is reported, along with the calculation of the resonant frequency that accounts for the radial and azimuthal wave numbers. The rest of the paper is organized as follows. Section II presents a detailed explanation of the mathematical principles and computational methods used to determine the field components. Additionally, it includes formulas to compute the resonant frequency of TE and EH modes for a sectoral CDRA. Section III presents the antenna's geometric structure and feeding mechanism, along with a parametric analysis of $EH_{110}$ and $TE_{210}$ modes. Section IV presents an analysis of the Specific Absorption Rate (SAR) for the proposed Quarter CDRA (QCDRA). It also includes the implementation of an Artificial Magnetic Conductor (AMC) surface to boost gain and manage SAR. Section V provides an in-depth analysis of the implementation and measurements of the prototype. Section VI concludes the article.

## Field components derivation

The existing work on CDRA for wearable applications is limited. The reason is typically the large size and thickness of DRAs. Low-profile sectoral CDRA comes in handy in this scenario. Instead of using a complete CDRA, a slice of it is employed by applying the appropriate boundary conditions. The existing literature is limited to TM modes and the resonant frequency relations available do not incorporate the azimuthal wavenumber in the calculation, thus providing a valid relation only for pure TM modes with no azimuthal variation. It is however important to understand that the interpretation of the field mode in sectoral CDRA differs from that of the complete CDRA. The azimuthal variation represented by the subscript $m$ is replaced with $v$[29]. The subscript $v$ represents a positive real number that is calculated based on boundary conditions and sector angle. The field components of $TE_{vnp}$ and $EH_{vnp}$ modes are derived in the cylindrical coordinate system $(r, \phi, z)$ in this section. The subscript $n$ represents the radial field variation and $p$ represents the axial variation. The longitudinal component of the electric field is zero for all $TE_{vnp}$ modes and approximately zero for $EH_{vnp}$ hybrid modes. The H field mainly contributes to the resonance ( $E_z = 0$ )[30]. The Helmholtz equation can thus be written in the form of a longitudinal H field (1)

$$\nabla^2 H_z(r, \phi, z) + k^2 H_z(r, \phi, z) = 0 \quad (1)$$

Dielectric resonator theory has been applied, and the Helmholtz equation has been solved using the separation of variables by assuming that the solution of (1) comprises the product of the independent solutions of their respective $r, \phi, z$ components, i.e.

$$H_z(r, \phi, z) = R(r) P(\phi) Z(z) \quad (2)$$

where $R(r)$ is the solution of the radial component of the differential equation, $P(\phi)$ is the solution of the azimuthal component of the equation and the $Z(z)$ is the solution of the axial component of the differential equation. Solving (1) by first taking the Laplacian and rearranging it to obtain three separate differential Eqs. (3, 4, 5)

$$\frac{d^2 P}{d\phi^2} + Pv^2 = 0 \quad (3)$$

where $v^2 = r^2 k_\phi^2$

$$r^2 \frac{d^2 R}{dr^2} + r \frac{dR}{dr} + R\left(k_r^2 - \frac{v^2}{r^2}\right) = 0 \quad (4)$$

$$\frac{d^2 Z}{dZ^2} + k_z^2 Z = 0 \tag{5}$$

The solution to the radial differential Eq. (3) will be in the form of a Bessel function of the first kind (6)

$$R(r) = C_v J_v(k_r r) \tag{6}$$

The solution for the azimuthal differential Eq. (4) will be in the sin and cosine form (7)

$$P(\phi) = A' \cos(v\phi) + B' \sin(v\phi) \tag{7}$$

whereas the solution to axial variation (5) will also be in the sin and cosine form Eq. (8). The axial component in Eq. (8) comprises two longitudinal waves traveling in both directions according to the cylindrical cavity resonator field distribution to accurately calculate the resonance[31]

$$Z(z) = E\cos(k_z z) + F\sin(k_z z) \tag{8}$$

Combining the solution of $R(r)$, $P(\phi)$ and $Z(z)$ in (2) and merging the constants generates the final solution of the Helmholtz Eq. (9)

$$H_z = [J_v(k_r r)][A\cos(v\phi) + B\sin(v\phi)][E\cos(k_z z) + F\sin(k_z z)] \tag{9}$$

By utilizing the solution to (9) and applying it to the general solution of the Maxwell equation for the cylindrical coordinate system[32], the field components for TE and EH modes have been derived (10, 11, 12, 13). The $E_z = 0$ for all equations since we are dealing with TE and EH modes.

$$E_r = \frac{-j}{k_r^2} \frac{\omega \mu}{r} [J_v(k_r r)](vB\cos(v\phi) - vA\sin(v\phi))(E\cos(k_z z) + F\sin(k_z z)) \tag{10}$$

$$E_\phi = \frac{j}{k_r^2} \omega \mu \left( k_r r [J_v'(k_r r)][A\cos(v\phi) + B\sin(v\phi)][E\cos(k_z z) + F\sin(k_z z)] \right) \tag{11}$$

$$H_r = \frac{-j}{k_r^2} k_z \left( k_r r [J_v'(k_r r)][A\cos(v\phi) + B\sin(v\phi)][E\cos(k_z z) + F\sin(k_z z)] \right) \tag{12}$$

$$H_\phi = \frac{-j}{k_r^2} \frac{k_z}{r} \left( [J_v(k_r r)](vB\cos(v\phi) - vA\sin(v\phi))[E\cos(k_z z) + F\sin(k_z z)] \right) \tag{13}$$

**Boundary conditions**
Since this work investigates QCDRA with metalized sector faces, the modified shape of CDRA has a Perfect Electric Conductor (PEC) assigned to inner surfaces ($\phi = 0$, $\phi = \pi/2$), and other boundaries are open. For simplicity, the open boundaries on top, bottom, and sides can be approximated as a Perfect Magnetic Conductor (PMC) due to the presence of high permittivity dielectric material. Now, to calculate the wavenumbers and the resonant frequency of the proposed QCDRA, appropriate boundary conditions are applied to Eq. (9).

*Top and bottom surfaces*
Since the bottom of the QCDRA is PMC, the normal component of the H field will be continuous, and its derivative will be zero $\frac{\partial H_z}{\partial z}\big|_{z=0} = 0$ or $\frac{\partial Z_z}{\partial z}\big|_{z=0} = 0$, thus we get $F = 0$ by solving the boundary condition at the bottom surface.

By applying the boundary condition on the top surface, $z = h$, $\frac{\partial H_z}{\partial z}\big|_{z=h} = 0$ or $\frac{\partial Z_z}{\partial z}\big|_{z=h} = 0$, taking the derivative of (9) and putting $F = 0$ we get $\sin(k_z h) = 0$

$$k_z = \frac{p\pi}{h} \tag{14}$$

*Outer circular surface*
Solving the PMC boundary condition at $r = a$, $H_\phi|_{r=a} = 0$, by equating (13) from field components to 0, we get $J_v(k_r r) = 0$ leading to the solution of $k_r$

$$k_r = \frac{X_{vn}}{r} \tag{15}$$

where $X_{vn}$ is the root of the Bessel equation of the first kind that makes $J_v(k_r r)$ zero.

*Inner faces*
Implementing PEC boundary conditions on $\phi = 0$ and $\phi = \pi/2$ the tangential component of E will be zero, that is, $E_r|_{\substack{\phi=0 \\ \phi=\pi/2}} = 0$, $E_r$ component is derived from the general solution of Maxwell's equation in (10).

For $\phi = 0$ surface, we have $B = 0$ and for $\phi = \pi/2$ we get $sin(v\phi) = 0$

$$v = \frac{m\pi}{\phi} \quad (16)$$

$$v = 2m \quad (17)$$

$$k_\phi = \frac{v}{r} = \frac{2m}{a} \quad (18)$$

To calculate the resonant frequency of the modes, the wave number equation is computed as

$$k^2 = k_r^2 + k_\phi^2 + k_z^2 \quad (19)$$

using the respective wave numbers of radial, azimuthal, and axial directions. Since the modes in Sectoral CDRA are not interpreted as a conventional CDRA, we must incorporate the azimuthal component in the calculation of the resonant frequency, which is given as

$$f_{vnp} = \frac{c}{2\pi\sqrt{\epsilon_r}}\sqrt{\left(\frac{X_{vn}}{a}\right)^2 + \left(\frac{v}{a}\right)^2 + \left(\frac{p\pi}{h}\right)^2} \quad (20)$$

## Antenna design and excitation

The QCDRA has been implemented using a 2.54-mm-thick Rogers TMM13 substrate ($h = 2.54$) of permittivity 12.85. A Coplanar Waveguide (CPW) feeding has been realized on a supporting substrate with a permittivity of 6.15 and a thickness of 1.6 mm. This feeding setup also enables the use of PEC boundary applications on the two inner faces. One of the inner faces has been coated with metal and connected to the ground plane of CPW, whereas the other face is coated with metal and connected to the feedline. It is important to mention that the ground has been eliminated beneath the CDRA to facilitate the generation of the TE mode. The $TE_{210}$ mode has been optimized through the radius of QCDRA ($a = 12$ mm). For QCDRA, the value of *m* in (17) is 1, thus $v = 2$[33]. The root of the Bessel function of the first kind $X_{21}$ is 5.1356. The resonant frequency of the mode, as determined by (20), is 6.12 GHz. The simulated resonant frequency shown in Fig. 1 for the mode can be seen at 5.75 GHz which is slightly less than predicted by the theoretical model. This difference is due to the feeding mechanism and the field interactions with the substrate and the ground plane around the CDRA. Figure 1 also displays the simulation design and the distribution of the electric field.

In the case of sectoral DRAs, the boundaries on the inner faces generate additional modes with resonant frequencies that are lower than the fundamental modes of a standard CDRA[34]. As follows from Eq. (20), the $EH_{110}$ mode ($a = 12mm$, $v = 1$, $n = 1$, $X_{11} = 3.8317$) can be excited in QCDRA with a resonant frequency of 4.39 GHz while maintaining the same dimension. The geometry of the feeding strip applied on the face of the QCDRA was modified to initiate the lower-order hybrid mode as shown in Fig. 2. The reflection coefficient of the modified feed is also shown in Fig. 2, where dual-mode resonance can be observed. A lower-order hybrid mode $EH_{110}$ is resonating around 4.74 GHz. The slight difference in the resonant frequency as compared to the theoretical model is caused by the underlying substrate and the metallic ground plane surrounding the CDRA.

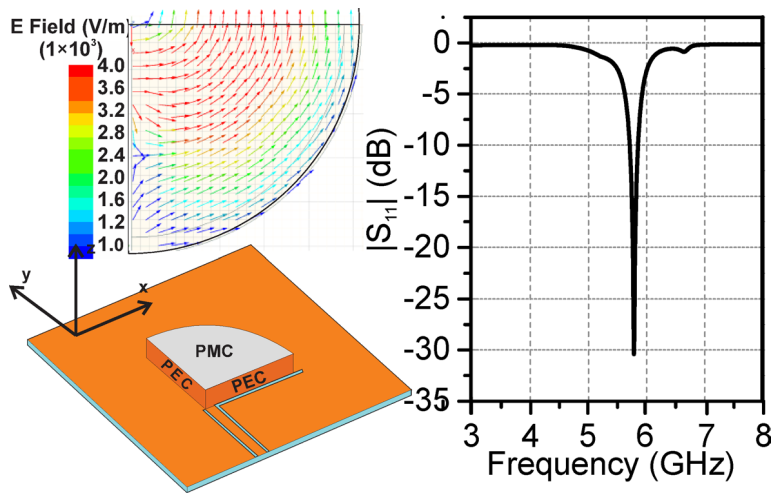

**Fig. 1**. Reflection coefficient, CPW feeding mechanism, and the electric field distribution for $TE_{210}$ mode in QCDRA.

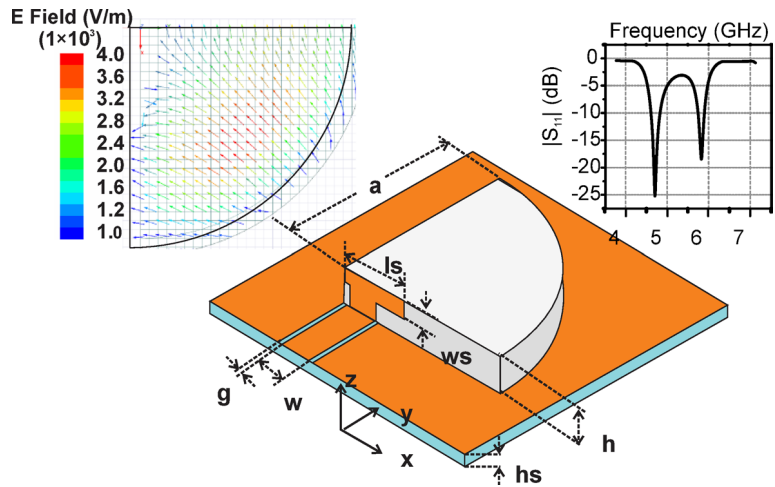

**Fig. 2**. Modified feed QCDRA geometry and the E-field vector plot of the $EH_{110}$ mode, the reflection coefficient of dual-mode QCDRA is shown in the inset.

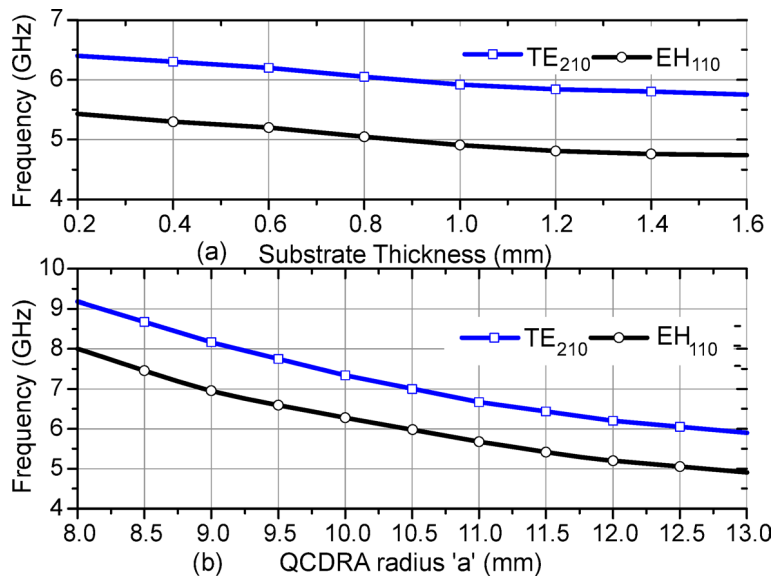

**Fig. 3**. Parametric analysis of (**a**) feeding substrate thickness (**b**) radius 'a' of QCDRA.

The parameters associated with the feeding strip do not affect the resonant frequency of the modes, since these modes are intrinsic to the QCDRA. These parameters just ensure the excitation of the desired modes by matching the field vectors with those of the desired modes. The resonant frequencies of the two modes are tunable through the radius of the QCDRA. Also, the substrate thickness underneath affects the two resonant modes; to investigate further the radius and the thickness of the underlying substrate have been swept over a wide range. Parametric analysis is shown in Fig. 3. Note that the radius of QCDRA changes the resonant frequency of the resonant modes. In particular, reducing the radius increases the resonant frequency which is also supported by the theoretical investigation in Eq. (20).

The required frequency band for the ISM application is around 5.8 GHz; thus the value of 12 mm radius has been adopted. Likewise, the substrate thickness also plays a role in controlling the resonant frequency. The thinner substrate has a higher resonant frequency of the modes and as we increase the thickness the resonant frequency is reduced as a thicker substrate has stronger field interaction as compared to the thinner ones. Following extensive optimization and practical reasons related to the substrate availability, the thickness of the substrate was set to 0.64 mm. Both resonant modes have been optimized at closer resonant frequencies, hence a wider operating band has been achieved as shown in Fig. 4. With the combination of two modes, the antenna offers a wide operating bandwidth of 25% from 4.9 GHz to 6.3 GHz. The optimized values for the whole structure are presented in Table 1.

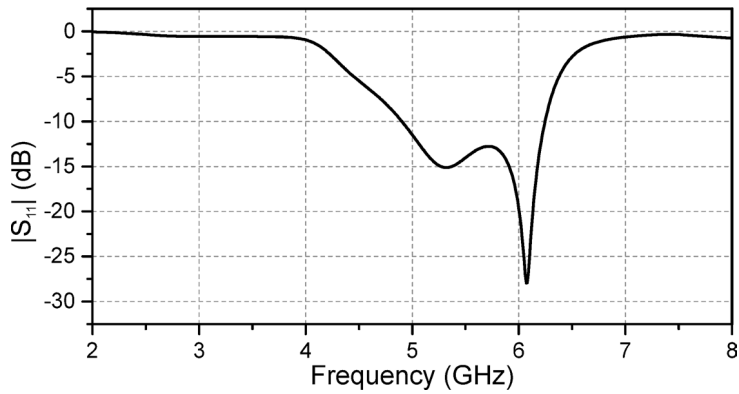

**Fig. 4**. Reflection coefficient of the optimized dual mode QCDRA.

| Parameter | g | w | l$_s$ | w$_s$ | h | h$_s$ | a |
|---|---|---|---|---|---|---|---|
| Value (mm) | 0.3 | 2.6 | 4.3 | 1 | 2.54 | 0.64 | 12 |

**Table 1**. Optimized parameters for QCDRA geometry.

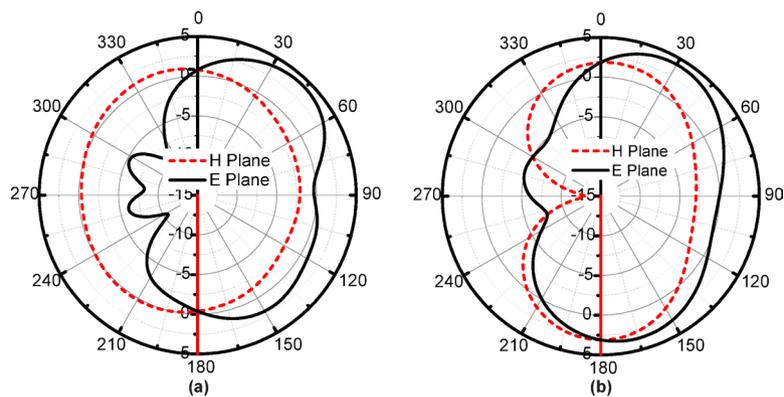

**Fig. 5**. E and H plane radiation patterns (a) $EH_{110}$ mode at 5.2 GHz (b) $TE_{210}$ mode at 6.1 GHz.

Once the design is optimized and a desirable operating band is achieved the radiation pattern has been simulated to verify the antenna gain patterns of the two modes. The simulated radiation patterns for both modes are shown in Fig. 5.

Both radiation patterns resemble that of a dipole radiator as anticipated with a large back radiation[35]. The radiation pattern for EH mode is predominately broadside; a major part of energy radiates towards the top and bottom. However, the radiation pattern shows less radiation towards the feeding section of antennas since the face is assigned PEC to launch the desired modes causing reflection to the radiating fields.

*SAR analysis and AMC introduction*

The proposed dual-band QCDRA can find applications in biomedical telemetry, and it is well established that exposing the human body to electromagnetic radiation increases the temperature of exposed parts. Consequently, extreme care is necessary in designing radiating devices that operate in close proximity to a human body. The level of exposure is quantified by the Specific Absorption Rate (SAR) as given in (21). SAR is the metric used to quantify the amount of energy absorbed by human tissues, calculated in W/kg[36].

$$SAR_{avg} = \frac{1}{V} \int \frac{\sigma(r) |E(r)|^2}{\rho(r)} dr \qquad (21)$$

In Eq. (21), $\sigma$ is the electrical conductivity of the medium, measured in $S/m$, $\rho$ is the mass density of tissues in $kg/m^3$, $|E|$ is the magnitude of the electric field in $V/m$, and $V$ is the volume of the sample taken at a radial distance $r$. The average SAR values of the proposed dual-mode antenna are simulated for a power of 1 W and the result is shown in Fig. 6. The SAR is highly localized and significantly elevated when the antenna is positioned 2 mm above the skin of the human arm.

Two primary standards are commonly referred to for the assessment and reporting of SAR in human tissue: the IEEE C95.1 standard and the European ECC/CEPT guidelines, which follow the recommendations of the International Commission on Non-Ionizing Radiation Protection (ICNIRP). According to IEEE C95.1 Standard, initially released in 1999, the average SAR limit for 1 g of tissue is fixed at 1.6 W/kg, whereas the following revisions in 2005 and 2019 introduced the average SAR limit for 10 g of tissues at 2 W/kg, also the peak SAR limit for 1 g tissue is set at 4 W/kg, where as the average SAR for 1 g tissue remains at 1.6 W/kg.

The second standard European ECC/CEPT Standards, based on initial guidelines in 1998, specify the average SAR limits to 1.6 W/kg for 1 g of tissue and 2 W/kg for 10 g of tissue[36,37].

The maximum allowed power is calculated using the relation given in (22) for the proposed CDRA.

$$P_{max} = P_{In} \times \left( \frac{SAR_{Limit}}{SAR_{achieved}} \right) \qquad (22)$$

Where:

*Pmax* is the maximum allowed power to be fed to the antenna to ensure the SAR remains in the prescribed limit.

*Pin* is the power input to the antenna to calculate the SAR, It is kept at 1 W.

$SAR_{achieved}$ is the SAR calculated at the input power *Pin*.

$SAR_{limit}$ is the maximum allowed SAR as per the followed standard.

The SAR value for a 1 W of input power is found to be 53.3 W/kg, and to ensure safety according to (22), the maximum allowed power will be only 37.5 mW.

This high SAR value in the proposed design is caused by the intense back radiation generated by the absence of a ground plane beneath the antenna. The ground plane cannot be added owing to TE and EH mode field distribution, a suitable solution for this issue is to use an Artificial Magnetic Conductor (AMC). The AMC serves as a Perfect Magnetic Conductor (PMC) boundary, maintaining the normal Hz component and reflecting the signal without any phase delay. This promotes the in-phase addition of the transmitted and reflected signals through constructive interference, resulting in higher gain[38]. This antenna design incorporates TE modes, and the addition of an AMC further enhances the support for these modes by maintaining normal H fields. As a result, the AMC can be positioned in close proximity to the QCDRA without affecting the field distribution. A compact AMC unit cell design has been created on Rogers 3010 substrate with a thickness of 1.27 mm for this application. The simulation setup of AMC with the periodic boundary conditions is illustrated in Fig. 7. The reflection phase response along with the impedance of the proposed AMC is also shown in Fig. 7. The design reflects the signal within ±90º phase variation in the band of 4.9 GHz to 5.9 GHz.

A 7×9 AMC surface has been created and positioned beneath the QCDRA. Multiple separation distances have been simulated to position the AMC as close as feasible without causing a notable effect on the QCDRA performance. The optimal separation distance of 5 mm (0.0867 $\lambda_0$) was determined, where the application of AMC does not have a substantial impact on the fields of QCDRA. The reflection coefficient of the QCDRA with AMC backing when worn on the arm of a human voxel is shown in Fig. 8.

It can be observed from Fig. 8 that the simulated bandwidth has decreased slightly on both higher and lower ends. The decrease in the operating band is mostly attributed to the limited bandwidth of the AMC sheet and its attachment to the human arm. In general, the QCDRA response adequately covers both the intended ISM band and the U-NII 5 GHz bands, without impacting the dual-mode performance. The radiation patterns for both modes have been improved significantly with the introduction of AMC backing as shown in Fig. 9. The back radiation is greatly reduced, and the broadside gain has risen to 8.2 dBi.

The AMC-backed QCDRA has been re-applied to the human arm to observe the revised value of SAR. As expected, the AMC addition has significantly reduced SAR. It dropped to 1.24 W/kg for 10 g of tissues when 1 W power is applied to it as shown in Fig. 10. The SAR of the proposed antenna system is now well within the approved limits, and it is distributed to extended areas on the arm with a lower power density as compared to non-AMC based QCDRA.

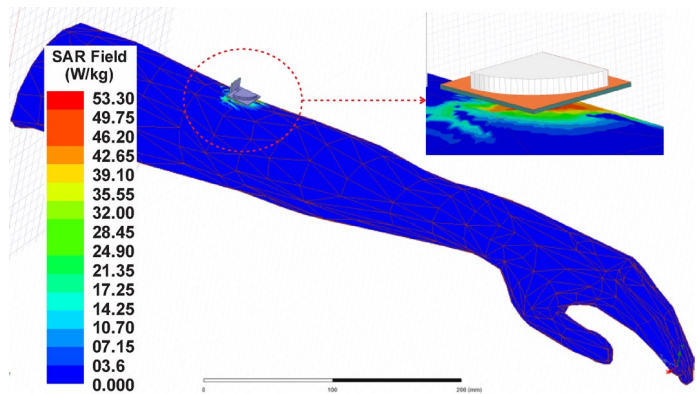

**Fig. 6.** The specific absorption rate of QCDRA with 1 W power for 10 g tissues.

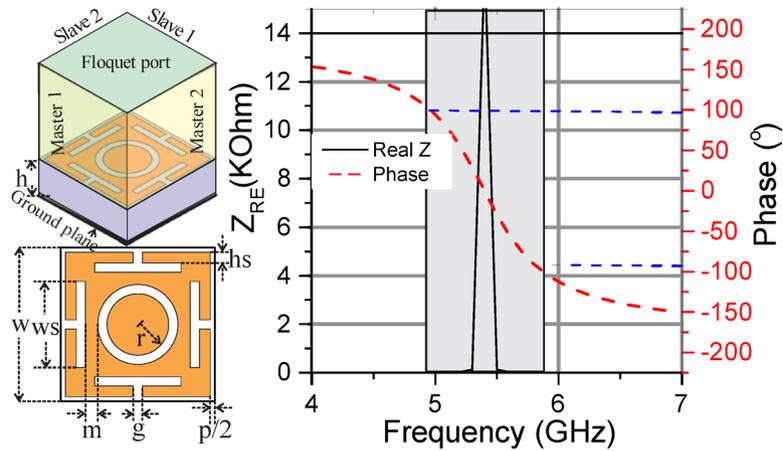

**Fig. 7.** AMC structure, simulation setup, phase and impedance response where $p/2 = 0.15$ mm, $g = 0.3$ mm, $h = 1.27$ mm, $h_s = 0.3$ mm, $w_s = 2.6$ mm, $w = 4.7$ mm, $m = 0.5$ mm, $r = 0.8$ mm.

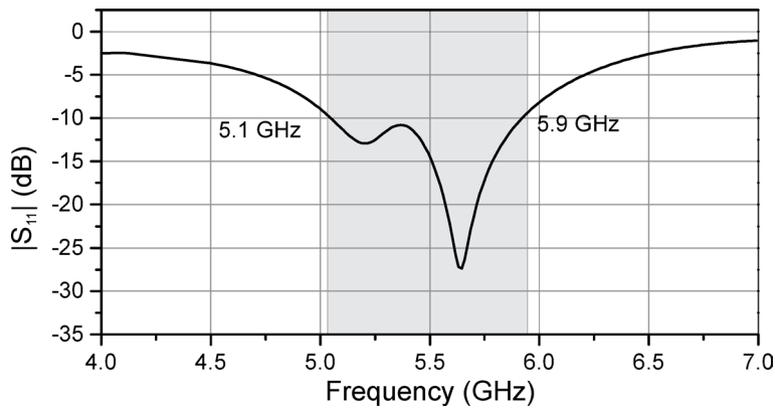

**Fig. 8.** Simulated reflection coefficient of QCDRA with AMC backing applied to a human arm.

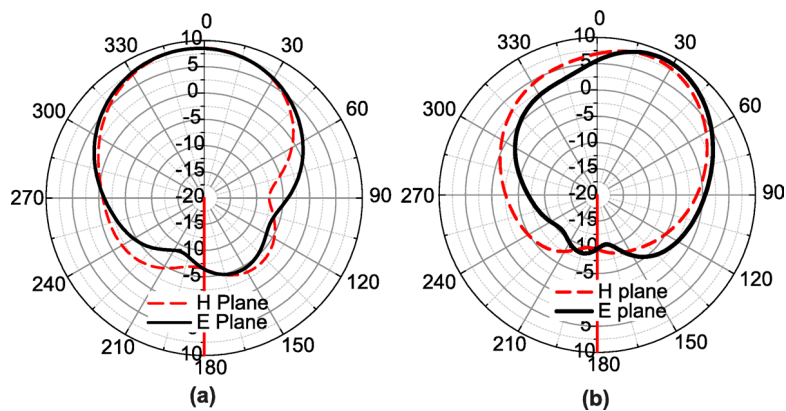

**Fig. 9.** Simulated radiation patterns with AMC (**a**) $EH_{110}$ mode @ 5.2 GHz (**b**) $TE_{210}$ mode @ 5.68 GHz.

A comparison of the proposed antenna without and with AMC is listed in Table 2. It can be observed that the peak gain has been increased significantly by the addition of AMC, which reflects the backward radiation and constructively adds to the broadside gain. The second major improvement is in the SAR, that reduced from above 50 W/kg to below 1.24 W/kg for $EH_{110}$ and 1.26 W/kg for $TE_{210}$ modes. The size, however, is increased, but the thickness is merely 0.16 $\lambda_o$ even after the addition of the AMC. The bandwidth is also reduced from 22.2 to 12.6% due to the intrinsic narrow-band nature of the metamaterial unit cell.

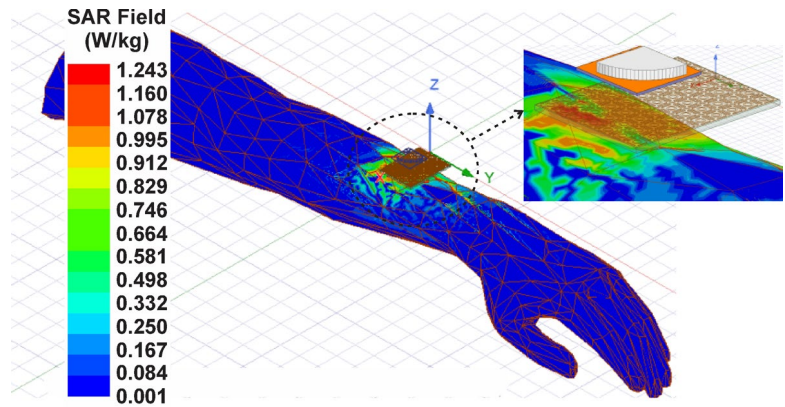

**Fig. 10**. SAR analysis of QCDRA with AMC backing for 1 W of applied power.

| | Antenna peak gain (dBi) | | Measured Antenna bandwidth | SAR for 10 g tissue (W/kg) | | Antenna Total efficiency | Size l×w×h |
|---|---|---|---|---|---|---|---|
| | $EH_{110}$ | $TE_{210}$ | (%) | $EH_{110}$ | $TE_{210}$ | (%) | ($\lambda_o$) |
| Without AMC | 4.1 | 4.4 | 22.2 | 53.3 | 54.8 | >86 | 0.32×0.34×0.05 |
| With AMC | 8.01 | 8.2 | 12.6 | 1.24 | 1.26 | >80.5 | 0.73×0.58×0.16 |

**Table 2**. Antenna comparison without and with AMC backing.

## Fabrication and measurement

The proposed QCDRA and AMC sheets were fabricated for testing. The QCDRA was implemented on a 2.54 mm thick Rogers TMM13 substrate of permittivity 12.85 and a loss tangent of 0.0019. A higher permittivity substrate is suitable for the implementation of DRA for better compactness. Rogers 3006 substrate with a thickness of 0.64 mm, permittivity of 6.15, and loss tangent of 0.002 has been used to act as the QCDRA hosting substrate. Any other substrate can also be used; this one is selected based on availability. The AMC sheet was realized on a 1.27 mm thick Rogers 3010 substrate with a permittivity of 10.2 and a loss tangent of 0.0022. A higher permittivity substrate for AMC ensures compact unit element size, thus it allows higher unit cell density thus the infinite periodicity of AMC surface is better approximated in a finite size.

The LPKF S103 milling machine was utilized to manufacture the prototype. The LPKF ProConduct conductive epoxy was applied to the faces of QCDRA. The grounded face was fully coated with epoxy and attached to the CPW ground. Meanwhile, the feeding strip was printed on the QCDRA face with the epoxy using a tiny stencil cut from a polyimide sheet. A 5 mm thick piece of Polymethacrylimide Foam ($\epsilon'_r$ = 1.043, tan$\delta$ = 0.0016) was utilized for the attachment of QCDRA and AMC. The fabricated prototype is shown in Fig. 11.

The reflection coefficient response of the prototype was measured using an Agilent N5242A PNA series network analyzer. The $S_{11}$ measurement of the QCDRA without AMC support is compared to the simulated response in Fig. 12. There is a minor decrease in bandwidth on the higher frequency end owing to fabrication imperfections, indicating a shift in the resonant frequency of the $TE_{210}$ mode. The observed bandwidth for QCDRA experiences a slight decrease from 25 to 22.2%.

The complete prototype of QCDRA with AMC backing has been measured while mounted on a human arm. The separation of 2 mm from the skin of the arm, mimicking a dress shirt, was realized with a foaming tape applied to the backside of the AMC sheet. The measured response is in good agreement with the simulated response as shown in Fig. 13. The measured response exhibits a modest decrease in bandwidth on the lower frequency side. The bandwidth ranges from 5.2 GHz to 5.9 GHz, encompassing the desired frequency band. Both resonators operate as expected and add to the overall range of performance.

The radiation response of the QCDRA with AMC backing was measured in a fully calibrated anechoic chamber. The prototype's response in both the E and H planes is presented in Fig. 14 for 5.2 GHz. Nulls are observed at approximately 90º and 270º. These nulls are more pronounced in the E plane and are primarily caused by diffraction and surface waves on the backing AMC. Increasing the number of cells in the AMC will reduce the back radiation, but this comes at the expense of the increased size. The broadside range of the radiation pattern around 0º is in excellent agreement with simulations and offers a maximum measured gain of 7.9 dBi.

Figure 15 illustrates the total efficiency of the stand-alone QCDRA and AMC-backed QCDRA, as well as when deployed on the human arm. The total efficiency of QCDRA without AMC exceeds 86% across the full frequency range of interest. However, when QCDRA is applied to the human arm, the total efficiency decreases considerably to below 70% due to the direct interaction of the lossy human arm with the radiating fields of the antenna. The introduction of AMC greatly enhanced the total efficiency and successfully restored it to a level exceeding 80%. Finally, when the AMC-backed QCDRA was applied to the human arm, a total efficiency of above 80% was achieved across most of the frequency range.

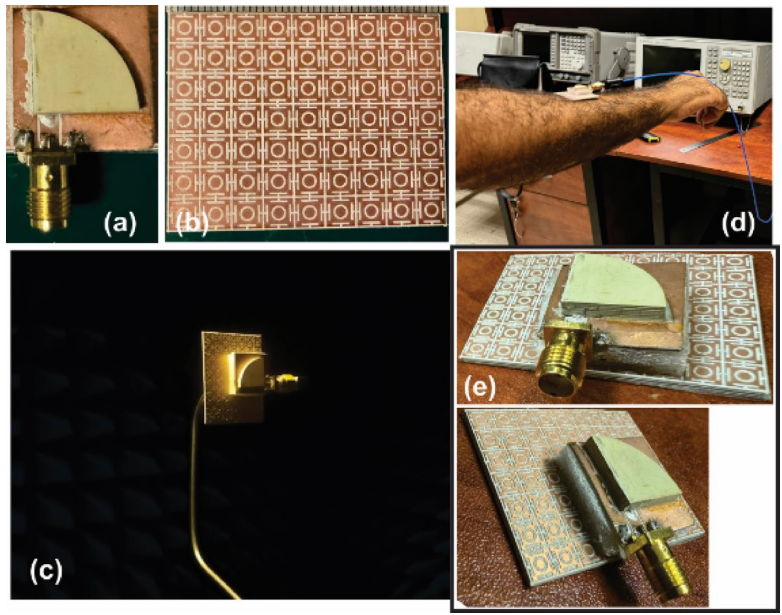

**Fig. 11**. Fabricated prototype, (**a**) QCDRA design, (**b**) AMC sheet of 7 × 9 cells, (**C**) radiation pattern measurement setup, (**d**) on arm testing, (**e**) completed prototype QCDRA with AMC.

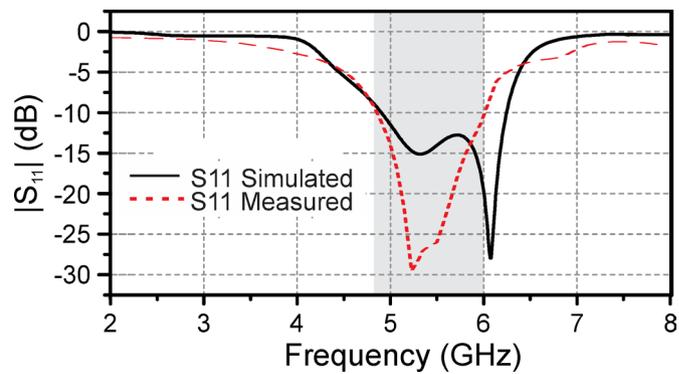

**Fig. 12**. Simulated and measured reflection coefficient of QCDRA without AMC.

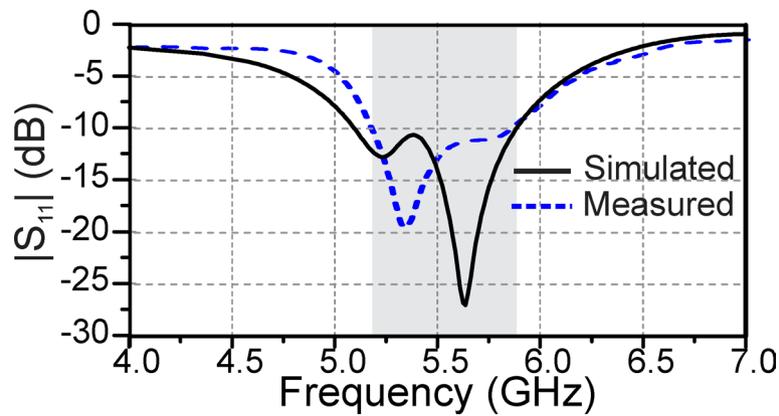

**Fig. 13**. Simulated and measured reflection coefficient of AMC-backed QCDRA applied on human arm.

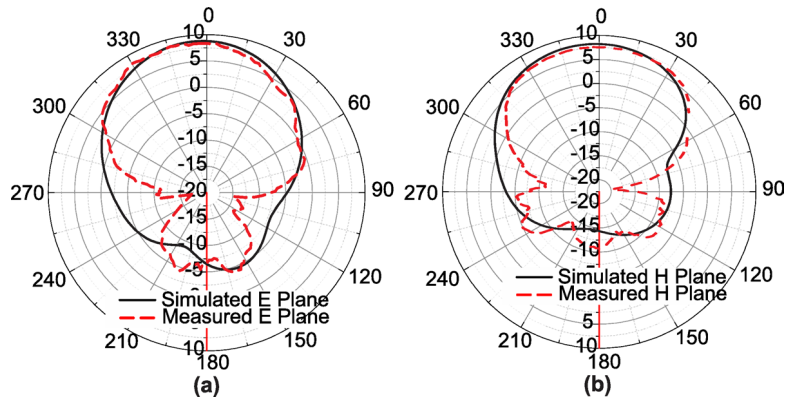

**Fig. 14.** Simulated and measured radiation patterns of AMC-backed QCDRA (**a**) E plane (**b**) H plane.

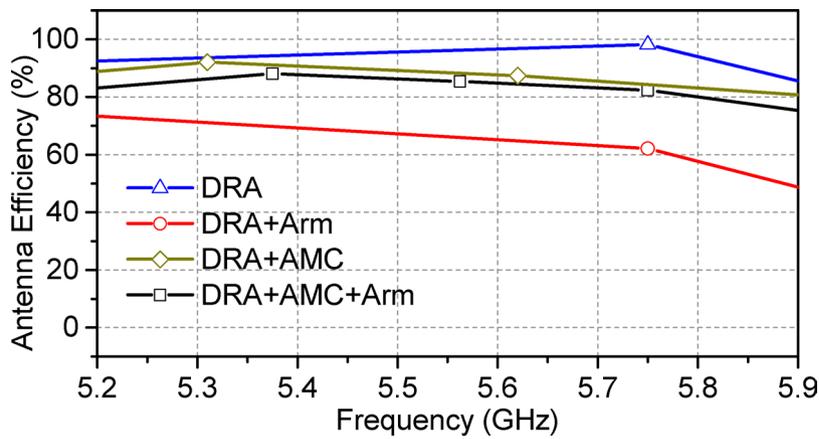

**Fig. 15.** Total Efficiency of QCDRA with AMC, without AMC with arm and without arm.

| Ref | Antenna Geometry | Frequency (GHz) | Measured Operational BW (%) | Antenna area $1 \times 10^{-2}$ $(\lambda_o)^2$ | Antenna Height $(\lambda_o)$ | Gain (dBi) |
|---|---|---|---|---|---|---|
| 16 | Cylindrical | 5.8 | 7 | 55.2 | 0.06 | 5.4 |
| 17 | Cylindrical | 2.4 | 4.95 | 27.0 | 0.1 | 1.25 |
| 18 | Rectangular | 6.7 | 16 | 110.1 | 0.33 | 7.5 |
| 19 | Conical | 3.3 | 42 | 43.5 | 0.13 | 2 |
| 20 | Cuboid | 7.8 | 9.6 | -- | 0.39 | 5 |
| 21 | Rectangular | 2.45 | 16 | 24.01 | 0.07 | 4.6 |
| 23 | Cylindrical | 5.47 | 7 | 9.96 | 0.21 | 3.5 |
| 24 | Cylindrical | 2.4/ 5.8 | Narrow band | 29.9 | 0.12 | -0.6/ 4.3 |
| 25 | Cylindrical | 2.4/ 5.8 | 5.6 | 8.47 | 0.13 | 2.2/ 3.6 |
| 26 | Cylindrical | 2.4/ 5.6 | Narrow band | 19.57 | 0.15 | -- |
| 27 | Cylindrical | 5.8 | 42 | 26.4 | 0.15 | 5 |
| This work with AMC | Cylindrical | 5.8 | 12.6 | 52.4 | 0.16 | 7.9 |

**Table 3.** Comparison with state-of-the-art wearable dielectric antennas.

To further underscore the significance of the proposed design, various state-of-the-art wearable antennas have been compared in Table 3 regarding their size, gain, and operational bandwidths. It can be observed that the proposed antenna has a moderate bandwidth, but it covers the ISM and 5 GHz U-NII bands. The proposed DRA area and height (including substrate) are compared with other DRAs. The QCDRA reported here has competitive dimensions as compared to the state-of-the-art. Also, it has the highest gain among the benchmark DRAs for wearable body area networks.

## Conclusion

This research proposed an innovative structure of a highly miniaturized dielectric resonator antenna. It demonstrates the first use of sectoral CDRA in biomedical telemetry paving the path for the adoption of CDRA in such applications. High antenna gain enables the device to establish the communication link at extended ranges, a capability not achievable with conventional patch antennas, thereby allowing the patient under observation to move freely from the gateway/base station. The antenna supports IEEE 802.11n and 802.11ac, enabling remote data access for biomedical sensors without requiring a gateway, thereby providing benefits over traditional gateway-based technologies by allowing direct internet connectivity with the help of a miniaturized MCU. The mathematical model for calculating all possible transverse and hybrid modes facilitates the application of QCDRA across various frequency bands, thereby promoting the rapid integration of this design into other biomedical technologies. The use of an AMC reduces the SAR by reflecting the radiation in-phase, thereby enhancing constructive interference. This enables the antenna to be designed with a separation of approximately 0.0867 $\lambda_0$ (5 mm), compared to a conventional PEC reflector, which requires a spacing of 0.25 $\lambda_0$ (quarter wavelength). Consequently, the AMC-based antenna achieves a thickness reduction of about 65% relative to the PEC-backed design. The SAR values remain within the prescribed limits even with 1 W of input power. The antenna has been implemented and evaluated for its performance when worn on the arm as well as while placed in an anechoic chamber.

## Data availability

The datasets used and/or analyzed during the current study available from the corresponding author on reasonable request.

### Acknowledgements
The authors thank Dassault Systemes, France, for making CST Microwave Studio available. This work is partially supported by the Nobelium Joining Gdansk Tech Research Community DEC-17/2021/IDUB/I.1 and by the Icelandic Research Fund Grant 2410297.

### Author contributions
Conceptualization, N.G. (Nayab Gogosh) and S.Kh. (Sohail Khalid); methodology, N.G. and S.Kh.; data generation, N.G. and B.T.M (Bilal Tariq Malik).; investigation, N.G. and S.Kh.; writing—original draft preparation, N.G., and B.T.M; writing—review and editing, S.K. (Slawomir Koziel) and S.Kh.; visualization, B.T.M and S.K.; supervision, S.K. and S.Kh.; project administration, S.K and B.T.M.

### Declarations

### Competing interests
The authors declare no competing interests.

### Additional information
**Correspondence** and requests for materials should be addressed to S.K.

**Reprints and permissions information** is available at www.nature.com/reprints.

**Publisher's note**  Springer Nature remains neutral with regard to jurisdictional claims in published maps and institutional affiliations.